\documentclass[journal,twoside,web]{IEEEtran}
\usepackage{cite}
\usepackage{amsmath,amssymb,amsfonts}
\usepackage{algorithmic}
\usepackage{graphicx}
\usepackage{textcomp}
\usepackage{subfig}
\usepackage{color}

\usepackage{bm}
\usepackage{xspace}
\newcommand{\bmath}[1]{\ensuremath{\bm{#1}}\xspace}

\newcommand{\x}{\bmath{x}}
\newcommand{\y}{\bmath{y}}
\newcommand{\z}{\bmath{z}}

\newcommand{\n}{\bmath{n}}

\newcommand{\tht}{\bmath{\theta}}

\newcommand{\etav}{\bmath{\eta}}

\newcommand{\A}{\bmath{A}}
\newcommand{\B}{\bmath{B}}

\newcommand{\F}{\bmath{F}}

\newcommand{\I}{\bmath{I}}

\newcommand{\Fc}{\bm{\mathcal{F}}}

\newcommand{\Lc}{\mathcal{L}}

\newcommand{\Oc}{\mathcal{O}}

\newcommand{\beq}{\begin{equation}}
\newcommand{\eeq}{\end{equation}}
\newcommand{\bea}{\begin{eqnarray}}
\newcommand{\eea}{\end{eqnarray}}
\newcommand{\ba}{\left(\!\!\begin{array}}
\newcommand{\ea}{\end{array}\!\!\right)}
\newcommand{\bc}{\begin{center}}
\newcommand{\ec}{\end{center}}

\newcommand{\LD}{\mathrm{LD}}
\newcommand{\ND}{\mathrm{ND}}

\def\BibTeX{{\rm B\kern-.05em{\sc i\kern-.025em b}\kern-.08em
		T\kern-.1667em\lower.7ex\hbox{E}\kern-.125emX}}

\ifCLASSINFOpdf
\else
\fi
\begin{document}
\title{Low-Dose CT Image Denoising Using Parallel-Clone Networks}
\author{Siqi Li and Guobao Wang
	\thanks{This work is supported in part by National Institutes of Health (NIH) under the grant no. R21EB027346 and K12CA138464.}
	\thanks{S. Q. Li and G. B. Wang are with the Department of Radiology, University of California Davis Medical Center, Sacramento, CA 95817, United States. (e-mail: sqlli@ucdavis.edu, gbwang@ucdavis.edu).}
}

\maketitle

\begin{abstract}
	Deep neural networks have a great potential to improve image denoising in low-dose computed tomography (LDCT). Popular ways to increase the network capacity include  adding more layers or repeating a modularized clone model in a sequence. In such sequential architectures, the noisy input image and end output image are commonly used only once in the training model, which however limits the overall learning performance. In this paper, we propose a parallel-clone neural network method that utilizes a modularized network model and exploits the benefit of parallel input, parallel-output loss, and clone-to-clone feature transfer. The proposed model keeps a similar or less number of unknown network weights as compared to conventional models but can accelerate the learning process significantly. 
	The method was evaluated using the Mayo LDCT dataset and compared with existing deep learning models. The results show that the use of parallel input, parallel-output loss, and clone-to-clone feature transfer all can contribute to an accelerated convergence of deep learning and lead to improved image quality in testing. The parallel-clone network has been demonstrated promising  for LDCT image denoising. 
\end{abstract}


\section{Introduction}
\label{sec:introduction}
X-ray computed tomography (CT) involves radiation exposure that may potentially increase the risk of genetic, cancerous, and other diseases in a patient \cite{b1}. To reduce these risks, low-dose CT (LDCT) imaging has become an attractive solution. 
However, LDCT with standard image reconstruction commonly results in high noise in the reconstructed images, which compromises the diagnostic performance. It is desirable to develop more advanced image processing methods to suppress the noise and improve the image quality in LDCT.

In general, there are two categories of image processing methods for improving LDCT. The first category is tomographic image reconstruction from projection data, including analytical reconstruction in combination with sinogram denoising (e.g., \cite{b5, b6, b7}),  model-based iterative reconstruction (e.g., \cite{Thibault07, Elbakri02, b10, b12, b41}), and deep learning based reconstruction (e.g., \cite{b17, b18, b19, b40, b42}). 
These methods have the advantage of exploiting the raw projection data more extensively but have the disadvantage that the access to raw CT projection data remains a resource barrier to many research groups. In contrast, post-reconstruction image denoising (e.g., \cite{b20, b21, b22}), the other category, directly deals with the reconstructed images, and are more widely accessible to the research community. A product, once developed, can also be  
more easily integrated into an existing clinical CT workflow.

\begin{figure*}[htp]
	\centering
	\includegraphics[trim=0cm 0cm 0cm 1cm, clip, width=6.2in]{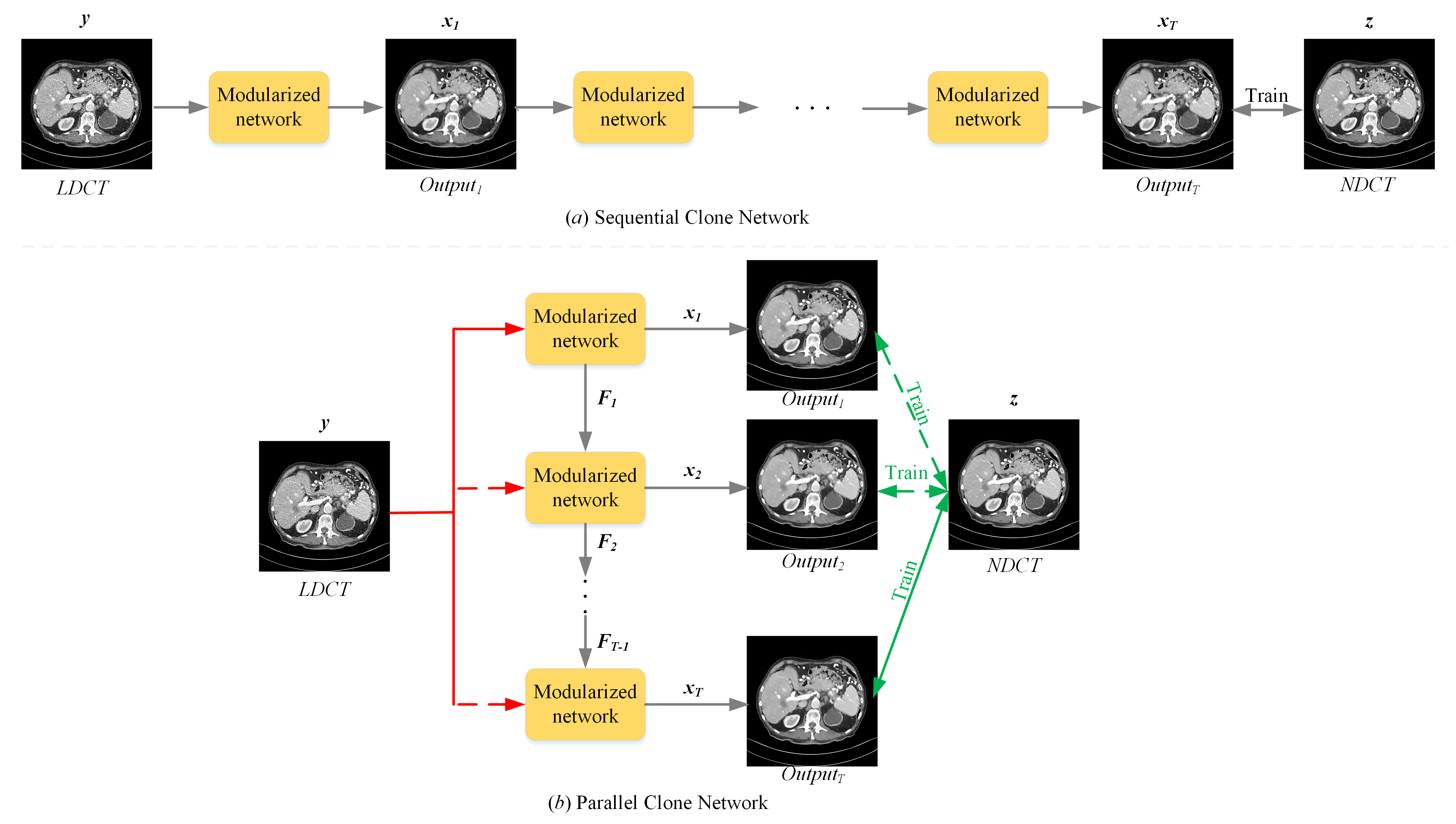}
	\caption{Conceptual illustration of (a) the sequential-clone network and (b) the proposed parallel-clone network. The sequential-clone model can be considered as a special example of the parallel-clone model if the LDCT input image is only fed to the first clone and only the output of the last clone is used in the training.
	}
	\label{fig:1}	
\end{figure*}

It has been demonstrated that deep-learning (DL) image denoising has a strong potential to improve LDCT \cite{b23, b24, b26, b27, b28, b29, b30}. The image quality by deep-learning image denoising can be equivalent to or even better than the state-of-the-art iterative CT reconstruction \cite{b29,b44}. A deep-learning model directly learns the end-to-end relationship between a noisy image and its clean reference image using, for example, deep neural networks (e.g., \cite{b24, b27, b28, b30}). Most existing network architectures improve the capacity of the neural network by adding more network layers. However, increased number of layers does not always improve the learning performance in practice\cite{b28}. A recent work of Shan \emph{et al.} \cite{b29} demonstrated an alternative solution that uses a modularized adaptive processing neural network (MAP-NN).  Instead of adding more new layers, MAP-NN repeatedly adds the same network module with shared weights (hence like ``clones") to increase the network depth and has demonstrated improved image quality for LDCT denoising. 

All the aforementioned deep learning methods for LDCT have a sequential-type layout as illustrated in Figure 1(a). From this perspective, the MAP-NN model is a sequential-clone network which uses multiple clones of the basic network module in a sequence of depth. Other earlier network models for LDCT denoising such as the residual encoder-decoder convolutional neural network (RED-CNN) \cite{b28} can be considered as a special case of the sequential-clone networks, in which only one clone of the basic network module (i.e., RED-CNN itself) is used. Increasing the number of clones from 1 deepens the network and has the potential to improve the training. 

Nevertheless, such sequential-clone architectures have two major limitations. First, the ability of forward propagating the raw image information is limited. The original noisy input image is used only once, i.e., in the first clone. As analyzed later in this paper, such a usage is very different from conventional model-based image denoising algorithms and can be less effective to propagate the useful information of the noisy input image into later clones.
Second, the network structure is also inefficient for back propagating the gradient of the loss function to earlier clones. The loss layer only utilizes the output image of the last clone and is far away from the earlier clones, which makes it difficult to back-propagate the gradient information of the loss to the earlier clones without causing a vanishing gradient.  As a result of these two limitations, the overall learning performance of the sequential-clone model for deep-learning image denoising can have been compromised.

In this paper, we propose a parallel-clone network method to overcome the limitations. The new model  exploits a parallel use of the noisy input image for the clones and incorporates the output image of all clones into the training loss function, also in parallel. The use of parallel input ensures an efficient forward propagation of useful information of the noisy input image into all the clones. The use of parallel output leads to an efficient back-propagation of the gradient of the loss function to the earlier clones. In addition, the proposed method also explores high-level feature transfer for the communications between the clones.  The proposed parallel-clone model is expected to bring substantial improvements over the existing sequential-clone framework for LDCT image denoising.


This paper is organized as follows. Section II introduces the backgroud materials that led to the development of the proposed method. Section III describes the detail of the proposed parallel-clone network model. Results of the training and testing on the Mayo CT Dataset are given in Section IV. Finally conclusions are drawn in Section V.

\section{Background}

\subsection{Model-based Image Denoising}

The forward model for traditional image denoising methods can be expressed as
\begin{equation}
\y=\A\x+\n,
\end{equation} 
where $\y$ and $\x$ denote the noisy CT image and the corresponding clean image to be estimated, respectively. $\A$ represents a degradation matrix and is equal to the identity matrix in this image denoising work. $\n$ represents the additive noise. 

The commonly used least-square image denoising problem is formulated as
\begin{equation}
\hat{\x} = \arg \min_{\x} \left\| {\y - \x} \right\|_2^2 + \lambda R(\x),
\label{eq-lsid}
\end{equation}
where the model consists of two components. The first is a data fidelity term and the second is a regularization item $R(\x)$ for exploiting the prior information with $\lambda$ the regularization parameter.  Iterative algorithms are commonly used to solve the optimization problem.

\subsection{End-to-End Deep Learning}

A learned model using neural networks predicts a denoised image $\x$ from the noisy image $\y$ using
\begin{equation}
\x\triangleq \psi (\y; \tht), 
\end{equation}
where $\psi$ denotes the end-to-end mapping from $\y$ to $\x$ with $\tht$ the neural network weights to be trained from available data sets.
A high-quality reference image $\z$ available from the training dataset may be equivalently expressed as
\beq
\z = \x + \etav ,
\eeq
where $\etav$ accounts for the difference between the prediction $\x$ and the truth $\z$. 
The mean squared error (MSE) between them is then defined by
\beq
\left\| \etav \right\|_2^2 = \left\| \z - \psi (\y; \tht) \right\|_2^2 \triangleq \Lc(\tht|\z, \y) .
\eeq
In LDCT image denoising, $\z$ corresponds to the normal-dose image $\I^{\ND}$ and $\y$ corresponds to the low-dose image $\I^{\LD}$. The training problem for deep-learning image denoising is then formulated as the following optimization if the MSE is used as the loss function:
\begin{equation}
\hat{\tht} = \arg \min_{\tht} \sum_{i=1}^{N} \Lc(\tht|\I^\ND_i, \I^\LD_i),
\end{equation}
where $i$ denotes the $i$th image pair of  low dose and normal dose in the training dataset. $N$ is the total number of training pairs. Once the model parameter set $\tht$ is trained, the final image estimate $\hat{\x}$ predicted from a noisy low-dose image $\I^\LD$ is obtained using $\hat{\x}=\psi (\I^\LD; \hat{\tht})$.

Examples of the neural network models for LDCT image denoising includes the RED-CNN \cite{b28}, wavelet residual network \cite{b24}, and so on.

\subsection{Sequential-Clone Neural Network}

To increase the depth of a neural network model, more layers can be added but with an increasing number of unknown model parameters. An alternative is to repeatedly use a network module. This concept has been explored in general deep-learning models such as the ResNet \cite{b36} and unrolled deep learning for image reconstruction (e.g., \cite{b18, b38}). A recent development of this concept for LDCT denoising is the modularized adaptive processing neural network (MAP-NN) \cite{b29} illustrated in Figure 1(a).  Mathematically the MAP-NN is expressed as
\beq
\x_T = \psi^T(\y;\tht),
\eeq
where $\psi^T(\y;\tht)$ denotes the repeated use of the modularized network $\psi(\cdot;\tht)$ for $T$ times:
\begin{equation}
\psi^T(\y;\tht) \triangleq \underbrace{\psi(\psi(\cdots\psi}_{T\;times}(\y; \tht);\cdots;\tht);\tht),
\end{equation}
which is equivalent to a sequence of ``clones": 
\beq
\x_1=\psi(\y;\tht), \x_2=\psi(\x_1; \tht), \cdots, \x_T=\psi(\x_{T-1}; \tht).
\label{eq-scn}
\eeq
Each clone is an individual denoiser but shares the same model structure $\psi$ and same parameters $\tht$ with other clones.   
The denoised image of a clone is the input of the next clone. If MSE is used, then the training loss for the sequential-clone network has the form of
\beq
\Lc_S(\tht|\z, \y) \triangleq\left\| \z - \psi^T(\y; \tht) \right\|_2^2.
\label{eq-scn-loss}
\eeq
If only one clone is used, i.e.,  $T=1$, the MAP-NN is then the same as traditional deep neural network models.


\section{Proposed Parallel-Clone Network}
Below we first describe the critical components that can be applied to the sequential-clone model individually. We then assemble them to form the proposed parallel-clone model as shown in Figure 1(b).

\subsection{Use of Coupled Input}

Residual mapping is popular in deep learning after the ResNet work \cite{b36}. The sequential model in Eq. (\ref{eq-scn}) can be equivalently rewritten as the following residual mapping format,
\beq
\x_t=\x_{t-1}+\phi(\x_{t-1};\tht),
\label{eq-scn2}
\eeq
where $\phi(\x_{t-1};\tht)$ denotes the residual mapping between the two adjacent clones and is mathematically equivalent to $\psi(\x_{t-1};\tht)-\x_{t-1}$\cite{b29} . Note that the noisy input image $\y$ appears only once in the sequence, i.e., in the first clone ($\x_0=\y$) but not in any subsequent clones. 

In comparison, conventional model-based image denoising commonly employs an optimization algorithm with the following iterative update:
\beq
\x^{(t)} = \x^{(t-1)}+\epsilon(\y,\x^{ (t-1)}),
\label{eq-it}
\eeq
where $\x^{(t)}$ denotes the image estimate at iteration $t$. $\epsilon$ represents the residual update determined by a specific algorithm. For example, the gradient descent algorithm for the least-square image denoising has the form of
\beq
\epsilon(\y,\x^{(t-1)}) =  \gamma [(\x^{(t-1)}-\y)+\lambda \nabla R(\x^{(t-1)})],
\eeq
where $\gamma$ is the step size and $\nabla R(\x)$ is the gradient of the regularization term. Another example is the alternating direction method of multipliers (ADMM) \cite{b31}, by which the iterative update can be described as
\begin{equation}
\eta(\y,\x^{(t-1)}) = \A^{(t)}\x^{(t - 1)} + \B^{(t)}\y,
\end{equation}
where $\A^{(t)}$ and $\B^{(t)}$ represent the updating matrices. 

One common feature of these model-based iterative denoising algorithms is that  the previous iterate $\x^{(t-1)}$ and the noisy input image $\y$ are coupled together to update the image estimate at next iteration $t$. This \textit{coupled input} is originated from the data fidelity term in the objective function Eq. (\ref{eq-lsid}). From the Bayesian perspective, the data fidelity carries useful information of the statistical distribution of measurements. We hypothesize the coupled use of $\x^{(t-1)}$ and $\y$ is more beneficial than using $\x^{(t-1)}$ alone.  

Considering each clone in the sequential model as an unrolled ``iteration", we can modify the sequential-clone model by including the noisy input image $\y$ into each clone,
\beq
\x_t= \x_{t-1}+\Phi(\y,\x_{t-1};\tht).
\label{eq-ci}
\eeq
Compared with Eq. (\ref{eq-scn2}), here we use a  different notation $\Phi(\cdot,\cdot;\tht)$ to denote the residual mapping now taking two inputs (i.e., $\x_{t-1}$ and $\y$) without significantly changing the structure of the modularized neural network $\phi$. The inputs $\y$ and $\x_{t-1}$ can be coupled using a  concatenation operation $Cat(\cdot,\cdot)$,
\beq
\F_t^{in} = Cat(\y,\x_{t-1}),
\eeq
which are then passed into the subsequent convolutional layers in the model $\Phi$. We expect this modification inspired from model-based image denoising can improve the residual mapping of each clone.

\subsection{Auxiliary Output Loss}

Compared to conventional model-based image denoising, deep learning has the advantage of end-to-end training. In the sequential-clone model, this is reflected in the training loss by comparing the output image of the last clone to the reference image, i.e., the normal-dose image in LDCT. 
Taking the MSE as an example, the training loss function for the sequential-clone network model is equivalent to 
\beq
\Lc_S(\tht|\z, \y) = \left\| \z - \x_T \right\|_2^2,
\label{eq-scn-loss}
\eeq
which only takes into account the output of the last clone $T$. 

A general challenge for deep learning is the vanishing gradient problem \cite{b39}, which holds true for the sequential-clone model. In a gradient-based algorithm, efficient back propagation of the gradient to the earlier clones in the sequence is challenging because the earlier clones are further away from the final layer of loss function than the later clones.

We note that in the sequential-clone network model, not only the last clone but also each of the earlier clones produces an auxiliary output image, which has not been utilized by the training process. In fact, auxiliary output has been utilized in previous works for the task of image recognition, e.g., GoogLeNet \cite{b37}. Hence we propose to  incorporate all the auxiliary output images into the loss function for the sequential-clone model. The MSE training loss is then of the form
\beq
\Lc_P(\tht|\z, \y) \triangleq \frac{1}{T}\sum_{t=1}^T \left\| \z - \x_t \right\|_2^2,
\label{eq-pcn-loss}
\eeq
in which the auxiliary output image of each clone contributes to the training loss in a parallel way. It becomes more straightforward to back-propagate the gradient information of the loss function from the output layer to the earlier clones. We expect the use of parallel auxiliary output loss can reduce the impact of the gradient vanishing problem.

\subsection{Brute-Force Residual Mapping}

The residual mapping function $\phi$ used in the sequential models mainly accounts for the difference between two adjacent clones,
\beq
\x_t-\x_{t-1}=\phi(\x_{t-1};\tht),
\eeq
which leads to the following form for the last clone,
\beq
\x_T=\y+\sum_{t=1}^T \phi(\x_{t-1};\tht).
\eeq
Substituting the above expression of $\x_T$ into the conventional loss function in Eq. (\ref{eq-scn-loss}), we have
\beq
\Lc_S(\tht|\z, \y) = \left\| (\z - \y)-\sum_{t=1}^T\phi(\x_{t-1};\tht) \right\|_2^2,
\label{eq-scn-loss2}
\eeq
which indicates that the total-residual image $(\z-\y)$ is approximated by a sum of the residual images  from all the $T$ clones. We call $\phi$ in this case incremental residual mapping. 

If the parallel auxiliary output loss in Eq. (\ref{eq-pcn-loss}) is used, then the training loss becomes to 
\beq
\Lc_P(\tht|\z, \y) = \frac{1}{T}\sum_{t=1}^T \left\| (\z - \y)-\sum_{\tau=1}^{t}\phi(\x_{\tau-1};\tht) \right\|_2^2,
\eeq
in which $\phi$ still represents an incremental residual mapping, though the accumulation is different in different clones.

Instead of using the mapping $\phi$ to represent the difference between two adjacent clones, i.e., $\x_t=\x_{t-1}+\phi(\x_{t-1};\tht)$, we employ a different residual mapping model in this work,
\beq
\x_t=\y+\phi(\x_{t-1};\tht),
\eeq
where the residual mapping $\phi$ is changed to reflect the difference between the output image $\x_t$ of clone $t$ and the noisy input image $\y$. Substituting the new expression into the parallel auxiliary output loss, we obtain
\beq
\Lc_P(\tht|\z, \y) = \frac{1}{T}\sum_{t=1}^T \left\| (\z - \y)-\phi(\x_{t-1};\tht) \right\|_2^2,
\eeq
which indicates  $\phi$ becomes to directly predict the total-residual image $(\z-\y)$ in each clone $t$. To be differentiated from the incremental residual mapping model, we call the new model as brute-force residual mapping in this paper.

In this work, we combine the brute-force residual mapping with the coupled input model $ \Phi(\y, \x_{t-1};\tht)$ to explore the benefit of parallel input as a part of the parallel-clone network model.

\begin{figure*}[htp]
	\centering
	\includegraphics[width=5.5in]{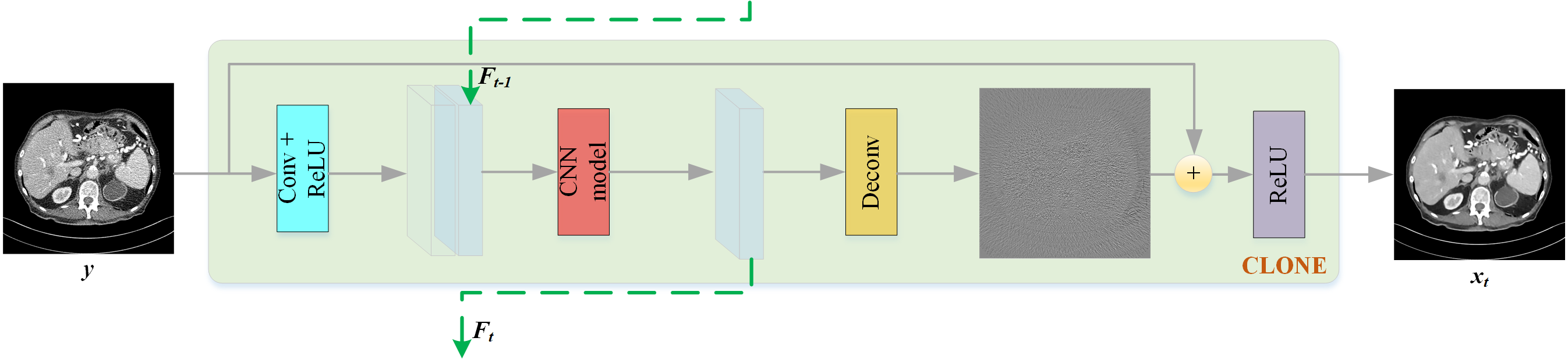}
	\caption{The architecture of a clone in the proposed parallel-clone network. It consists of a low-level feature extraction layer, a multilayer CNN module, and an image recovery module with residual mapping. The high-level feature set from an earlier clone is transferred into the next clone and combined with the low-level feature to form the input of the CNN module. }
	\label{fig:2}	
\end{figure*}

\subsection{Clone-to-Clone Feature Transfer}

In the sequential-clone model in figure 1(a), adjacent clones (e.g., $t$ and $t-1$) are connected using the intermediate denoised image. Other than the output image $\x_{t-1}$, additional high-level features also exist from the clone $t-1$ and can be transferred to clone $t$. We hypothesize the transfer of intermediate high-level feature can be more useful than just transferring the output image. Thus we use an a more general expression for the model of clone $t$:
\beq
\x_t= \x_{t-1} +\Phi(\y,\F_{t-1};\tht).
\eeq
where $\F_{t-1}$, the transferred information from clone $t-1$, can be the output image $\x_{t-1}$ or a high-level feature set.

In order to jointly use $\y$ and $\F_{t-1}$ if the latter represents high-level features, the clone model first extracts a feature set $\F_0$ from $\y$ using
\beq
\F_0 = \Fc_l(\y),
\eeq
where $\Fc_l(\cdot)$ denotes a low-level feature extraction operation and is composed of a convolutional layer $Conv()$ and a rectified linear unit (ReLU) layer $ReLU()$ in this work. The extracted feature set $\F_0$ matches the dimension of $\F_{t-1}$ but is more focused on the low-level information of the input image, such as edge and corner \cite{b32}. $\F_0$ and $\F_{t-1}$ can be then concatenated, 
\beq
\F_t^{in} = Cat(\F_0,\F_{t-1}),
\label{eq-hfin}
\eeq
to form the input for the subsequent convolutional layers.

\subsection{Combined Parallel-Clone Network Model}

Combing all the aforementioned components together, we obtain a general expression for the proposed parallel-clone network model,
\beq
\x_t=\y+\Phi(\y,\F_{t-1};\tht).
\eeq
The objective function for the corresponding optimization problem is then defined by
\beq
\Lc_P(\tht|\z, \y) = \frac{1}{T}\sum_{t=1}^T \left\| \z - \y - \Phi(\y,\F_{t-1}; \tht) \right\|_2^2.
\label{eq-pcn-loss-final}
\eeq
A graphical illustration of the parallel-clone model is provided in Fig. 1(b). 

This parallel-clone network model has three unique features: (1) parallel input, (2) parallel-output loss, and (3) clone-to-clone feature transfer. The parallel input feeds the noisy input image $\y$ to all the clones in parallel to enable the brute-force residual mapping and the use of coupled input, both improving the forward propagation of image information to the loss layer. The parallel-output loss incorporates all the auxiliary outputs into the training loss, which can improve the backpropagation of the gradient for the earlier clones. The clone-to-clone feature transfer connects adjacent clones to allow deeper learning.

The parallel-clone model is equivalent to the sequential-clone model if the input image is only fed to the first clone and only the output of the last clone is used in the training.

\subsection{Example Architecture and Implementation}

An example of the specific architecture of the proposed parallel-clone network is shown in figure 2. The model consists of  three modules in each clone:  a low-level representation extraction module $\Fc_l(\cdot)$ to obtain the low-level feature set $\F_0$, a high-level representation learning module $\Fc_h(\cdot)$ to obtain the high-level feature $\F_t$, and an image recovery module $\Oc(\cdot)$ to get the output image $\x_t$. Different clones have the same model structure with shared weights.

The image recovery module $\Oc(\cdot)$ is implemented using a deconvolution layer $Deconv(\cdot)$, a residual connection, and a $ReLU(\cdot)$:
\begin{equation}
\x_t = \Oc(\y, \F_t) = ReLU\big(\y + Deconv(\F_t)\big),
\end{equation}
where $Deconv(\cdot)$ outputs the residual image from the high-level feature set $\F_t$. 

The high-level representation learning $\Fc_h(\cdot)$ for clone $t$ is implemented by a multi-layer convolutional neural net (CNN),
\beq
\F_t=\Fc_h(\F_t^{in}),
\eeq
where its input $\F_t^{in}$ is a concatenation of the low-level feature $\F_0$ and the high-level feature $\F_{t-1}$ transferred from the clone $t-1$, see Eq. (\ref{eq-hfin}).

\begin{figure}
	\centering
	\includegraphics[width=3.5in]{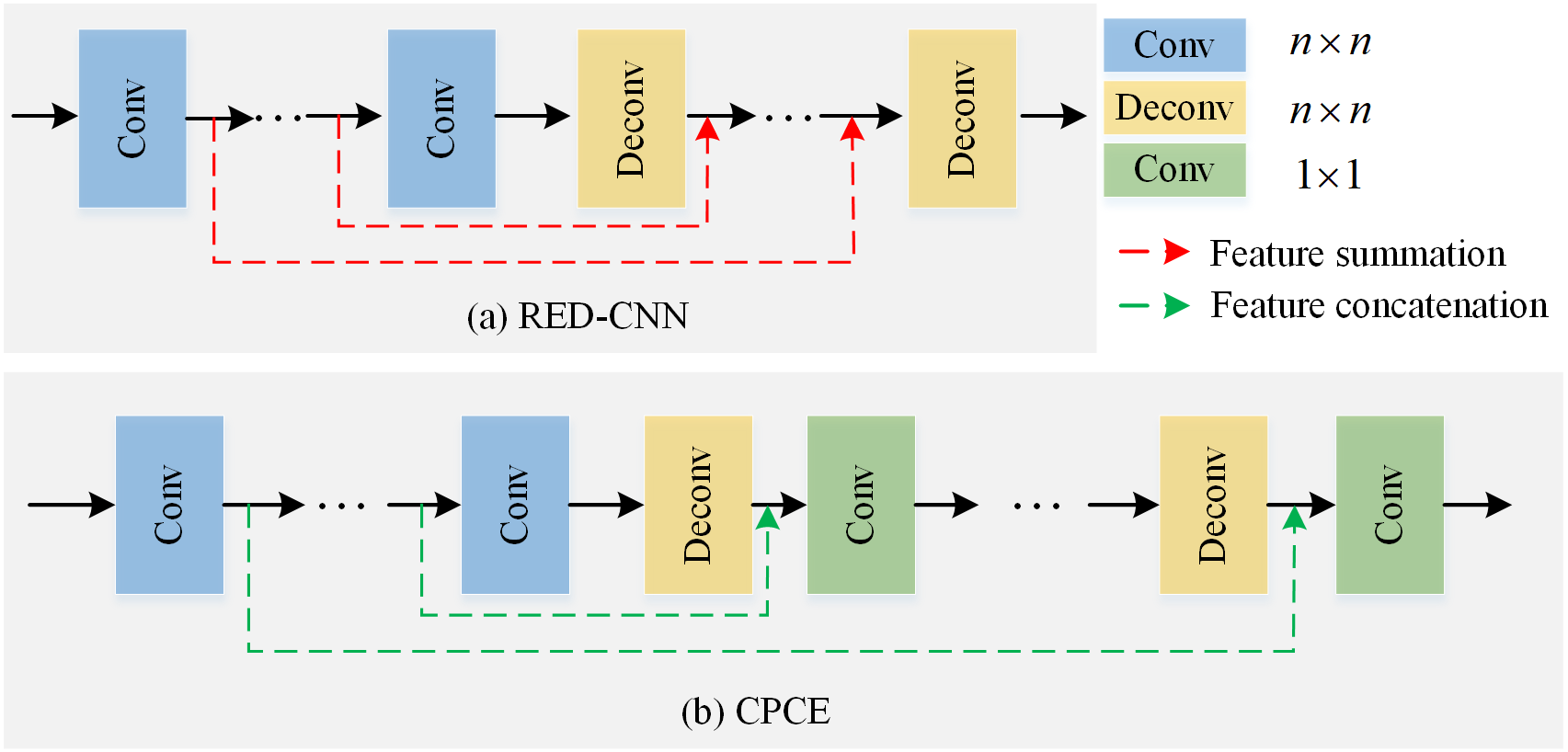}
	\caption{Example of basic CNN module that can be used for the clone network models. (a) RED-CNN \cite{b28}, (b) CPCE \cite{b29}.}
	\label{fig:3}	
\end{figure}

In theory, any deep-learning model can be used as the CNN module for the clone networks.  Fig. 3 shows two examples adapted from the RED-CNN model \cite{b28} and the CPCE model \cite{b29}, both following an encoder-decoder architecture. Details of the models are referred to the original papers of these models. The encoders consist of a series of $Conv()$ followed by $ReLU()$ to suppress image noise and artifacts from low-level to high-level step by step. 
In the decoders, a series of $Deconv()$ are used with residual mapping to recover the structural details. As $Conv()$ and $Deconv()$ are symmetric in the models, the number of $Deconv()$ is the same as the number of $Conv()$. The deconvolution in combination with symmetric shortcut connections are beneficial for detail preservation \cite{b28}.

\begin{figure*}[t]
	\centering
	\subfloat[]{\includegraphics[trim=0cm 0cm 0cm 2cm, width=3in]{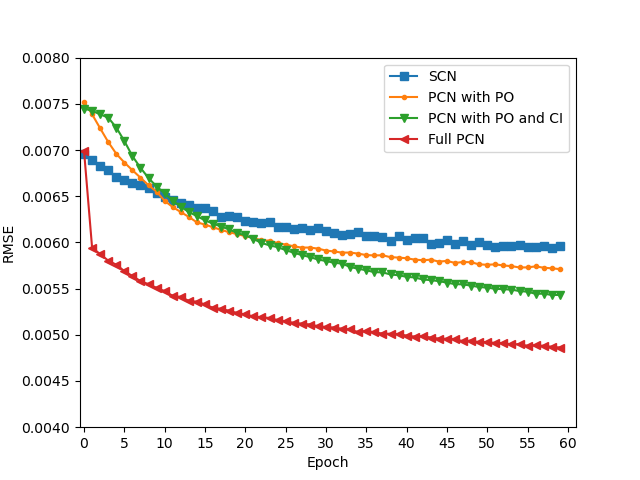}
		\label{fig_2_case}}
	\hfil
	\subfloat[]{\includegraphics[trim=0cm 0 0cm 2cm, width=3in]{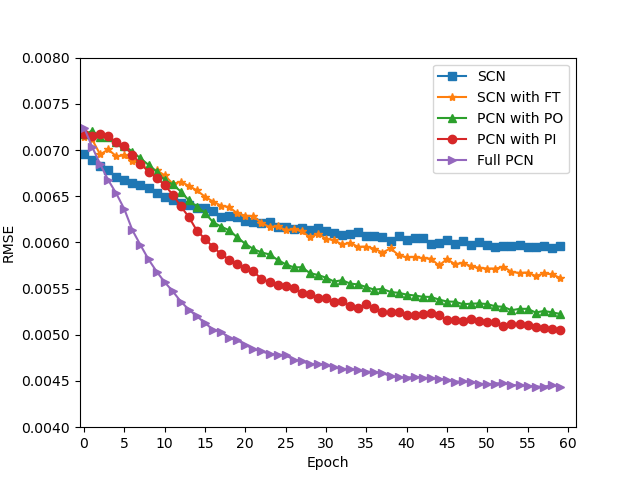}
		\label{fig_2_case}}
	\caption{~Plots of the RMSE convergence curve of the training data for different models based on (a) clone-to-clone image transfer and (b) clone-to-clone feature transfer. PO, CI, FT, and PI denote parallel input, coupled input, feature transfer, and parallel output,  respectively. The metrics were evaluated based on the training image patches.}
	\label{fig:4}	
	\vspace{-10pt}
\end{figure*}

\begin{figure*}
	\centering
	\subfloat[]{\includegraphics[trim=0cm 0cm 0cm 2cm, width=3in]{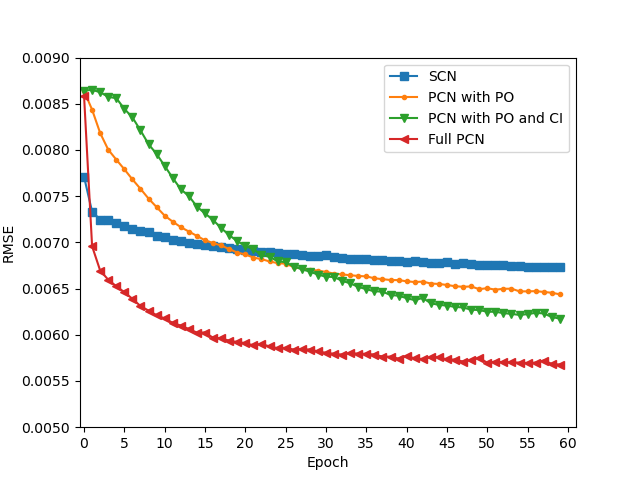}
		\label{fig_2_case}}
	\hfil
	\subfloat[]{\includegraphics[trim=0cm 0cm 0cm 1cm, width=3in]{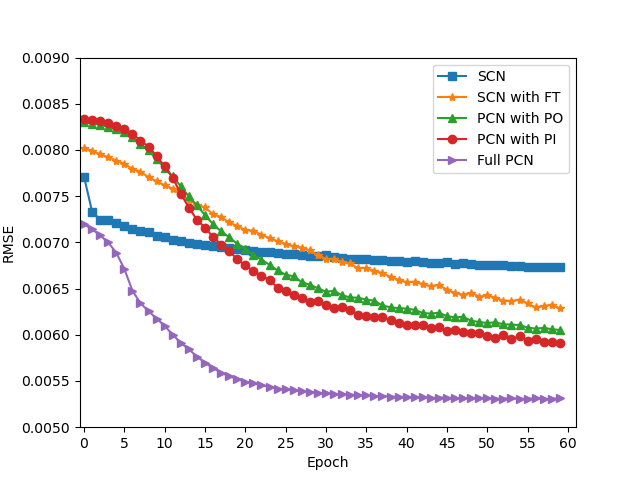}
		\label{fig_2_case}}
	\caption{~Plots of the RMSE convergence of the testing image data for different models based on  (a) clone-to-clone image transfer and (b) clone-to-clone feature transfer.  The abbreviations are the same with Fig. 4. }
	\label{fig:5}	
\end{figure*}

\section{Experiments and Results} 

\subsection{Clinical CT Dataset and Implementation}

The 2016 NIH-AAPM-Mayo Clinic Low Dose CT Grand Challenge dataset \cite{b33} was used for evaluating the proposed parallel-clone network and other models. The dataset includes the normal-dose abdominal CT scans and synthetic quarter-dose CT scans of ten patients. Each scan consists of about 210 to 340 transverse image slices, each with a matrix size of 512$\times$512 pixels. Nine out of ten were randomly selected and used for training and the remaining one was used for testing. The process was repeated for three times.

For each training, ten image slices were randomly selected from each of the nine patients to generate image patches of size 55$\times$55 with an interval of 4 pixels. The resulting total number of image patches used for training was about 1.6 million. For testing, the trained model was directly applied to the full image slices from the testing patient scan.

\begin{figure*}[htp]
	\centering
	\includegraphics[width=5.5in]{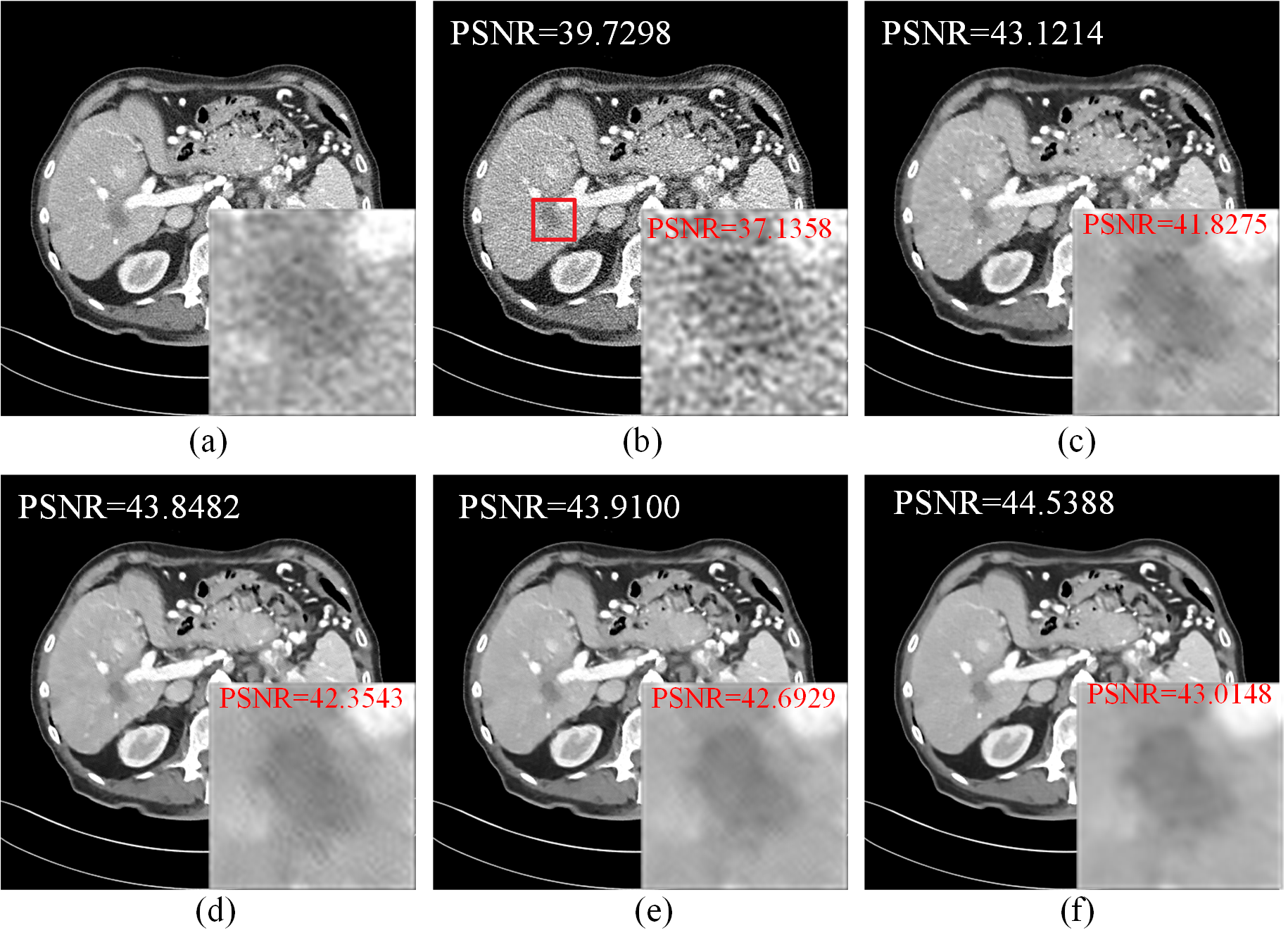}
	\caption{~Comparison of different models for denoising a testing image. (a) Norma-dose CT, (b) LDCT,  (c) SCN, (d) PCN with parallel-out loss only, (e) PCN with parallel-input only, and (f) PCN with all components integrated. 
	}
	\label{fig:6}	
\end{figure*}

We used the Adam optimization method \cite{b34} to train different network models with a mini-batch of 128 patches in each iteration. Sixty epochs were run. The initial learning rate was set to $10^{-4}$ and slowly decreased to $10^{-5}$. The number of convolutional kernels in each layer was 48 except for the last layer, which has only one layer. The kernel size of all layers was set to 3$\times$3 with a convolutional stride of 1 and no padding. All the networks were implemented using  PyTorch on a PC with an Intel i9-9920X CPU with 64GB RAM and a NVIDIA GeForce RTX 2080 Ti GPU.

\subsection{Approach of Comparison}

We first conducted an ablation study to demonstrate the improvement from the use of parallel input, parallel-out loss, and clone-to-clone feature transfer by using the sequential-clone network (SCN) model as the baseline. The RED-CNN model was used as the basic CNN module for both the SCN model and parallel-clone network (PCN) model. The output image of the last clone was used as the final output of each model unless specified otherwise. We also investigated the effect of hyperparameters in the PCN model, such as the number of layers for the CNN module and the number of clones.

The PCN model was further compared with three popular DL-based denoising methods: the denoising convolutional neural network (DnCNN) \cite{b35}, RED-CNN \cite{b28} and MAP-NN \cite{b29}.  DnCNN is one of the representative DL models for general image denoising. RED-CNN is a typical example specifically developed for LDCT denoising. The MAP-NN reflects the most recent example of a sequential-clone model for LDCT image denoising. 

Note that the DnCNN, RED-CNN, and proposed PCN models were trained using the MSE-based loss.  While the original MAP-NN was trained using an advanced loss function that combines the basic MSE loss with  perceptual losses, we also trained another MAP-NN using the MSE loss only. 


\subsection{Evaluation Metrics}

Three common image quality metrics, including the root mean square error (RMSE), peak signal to noise ratio (PSNR),  and structural similarity index measure (SSIM) \cite{b17}, were used to assess the training convergence and testing image quality.
\begin{equation}
\mathrm{RMSE} = \sqrt{\frac{1}{n_i}\sum_{i=1}^{n_i} (z_i-x_i)^2},
\end{equation}
\begin{equation}
\mathrm{PSNR} = 20{\log _{10}}\frac{I_{\max}}{\mathrm{RMSE}},
\end{equation}
\begin{equation}
\mathrm{SSIM} = \frac{{(2{\mu _z}{\mu _{x}} + {c_1})(2{\delta _{z,x}} + {c_2})}}{{({\mu _z}^2 + {\mu _{x}}^2 + {c_1})({\delta _z}^2 + {\delta _{x}}^2 + {c_2})}},
\end{equation}
where $\z$ and $\hat \x$ are the normal-dose image and predicted low-dose image with $n_i$ the total number of pixels in the region for quality evaluation. $\mu$ and $\delta$ denote the mean value and standard deviation inside a sliding window. $\delta_{z,x}$ respresents the covariance between $\z$ and $\x$. $I_{\max}$ is the maximum pixel value. $c_1$ and $c_2$ are two SSIM parameters defined as $(0.01 \times I_{\max})^2$ and $(0.03 \times I_{\max})^2$, respectively.

\subsection{Comparison Between Sequential and Parallel Models}

Fig. 4(a) shows the plots of the training convergence of RMSE as a function of epoch number for different clone models based on the clone-to-clone image transfer. The RMSE was calculated based on the training image patches and averaged over the the 3 times experiments. For each DL model, the number of clones was four and the number of layers in the basic RED-CNN module was ten. Compared with SCN, the use of parallel output (PO) and coupled input (CI) incrementally improved the RMSE. Further combination with the brute-force residual mapping model, which leads to the full PCN, achieved a significant acceleration of the training convergence as compared to the SCN model. 

Fig. 4(b) further shows the comparison based on the clone-to-clone feature transfer (FT). Replacing the image transfer with FT can improve the RMSE, though not for earlier epochs. On top of that, the use of parallel-output loss further improved the convergence. An even larger improvement was obtained with the use of parallel input, which  includes both the brute-force residual mapping model and coupled input. The most significant improvement came from the full PCN model which integrates all the three critical modifications (FT, PO, and PI) in the model. The convergence rate of the PCN was dramatically faster than the SCN. The PCN only took about 5 epochs to reach a similar RMSE value as the SCN at 60 epochs, suggesting a speed-up factor of about ten. 

Fig. 5 shows the results from the evaluation on the testing data. Note that here the image quality was evaluated on the whole image slices. The relationships between different models are consistent with the results of training convergence shown in Fig. 4. 

\begin{figure}[t]
	\centering
	\includegraphics[trim=0cm 0cm 0cm 1.5cm, width=3.0 in]{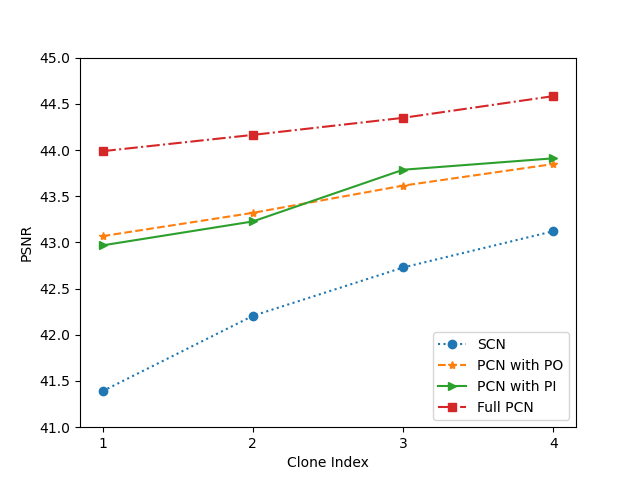}
	\caption{~Clone-wise comparison of PSNR between the SCN and different PCN models with four clones for denoising a testing image.}
	\label{fig:pcn-scn-psnr}	
\end{figure}

Fig. 6 shows the denoised results of a specific testing image by the SCN model and PCN models with 60 epochs. All the PCN models were implemented with the clone-to-clone feature transfer. For better display, the region of a liver metastasis was magnified in each image. Compared to the normal-dose reference image, the SCN reduced the noise but suffered from artifacts. The use of parallel-output loss or parallel input alone improved the image denoising according to quantitative PSNR.
The full PCN model which integrates the parallel input, parallel-output loss, and clone-to-clone feature transfer together achieved the best result in terms of quantitative PSNR and visual quality. These results are further confirmed in Fig. \ref{fig:pcn-scn-psnr} which shows a clone-wise comparison of PSNR for the SCN and different PCN models that consist of four clones. 

Table I summarizes the results of different quality metrics (PSNR, SSIM, and RMSE) from the testing dataset. The mean and standard deviation (SD) were shown for each metric. This comparison further confirmed the improvement of the full PCN model and its individual components (parallel-output loss, parallel input, and clone-to-clone feature transfer) as compared to the baseline SCN model.

\begin{table}[t]
	\centering 
	\caption{~Comparison of image PSNR (mean$\pm$SD) for SCN and PCN with different options.}
	\begin{tabular*}{3.5in}{@{\extracolsep{\fill}}lccc}
		\hline
		Methods	&PSNR &SSIM&RMSE \\
		\hline
		SCN& 43.5824$\pm$1.1476 & 0.9624$\pm$0.0054&0.0069$\pm$0.0011\\
		PCN with PO & 44.0179$\pm$1.2873& 0.9705$\pm$0.0052\ &0.0061$\pm$0.0007\\
		PCN with PI& 44.0268$\pm$1.3675&0.9718$\pm$0.0056&0.0060$\pm$0.0008\\
		Full PCN & \textbf{45.4235$\pm$1.1394}&\textbf{0.9846$\pm$0.0047}&\textbf{0.0054$\pm$0.0006}\\
		\hline
	\end{tabular*}
	\label{tab:my_label}
\end{table}

\subsection{Effect of Clone Settings}

\begin{figure}
	\centering
	\subfloat[]{\includegraphics[trim=0cm 0cm 0cm 1.3cm, clip,width=2.8in]{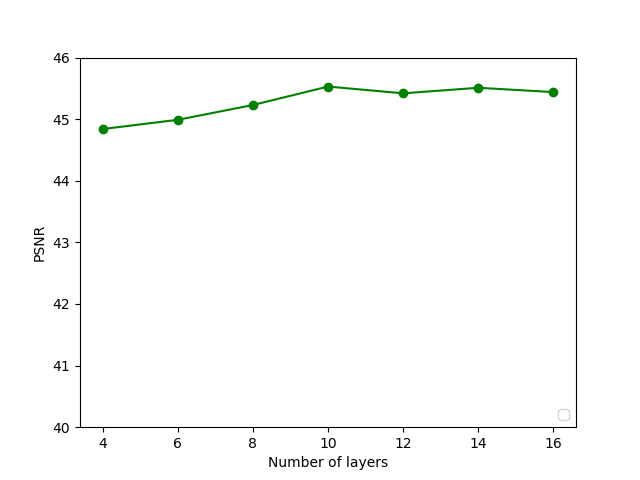}
		\label{fig_2_case}}
	\hfil
	\subfloat[]{\includegraphics[trim=0cm 0cm 0cm 0.7cm, width=2.8in]{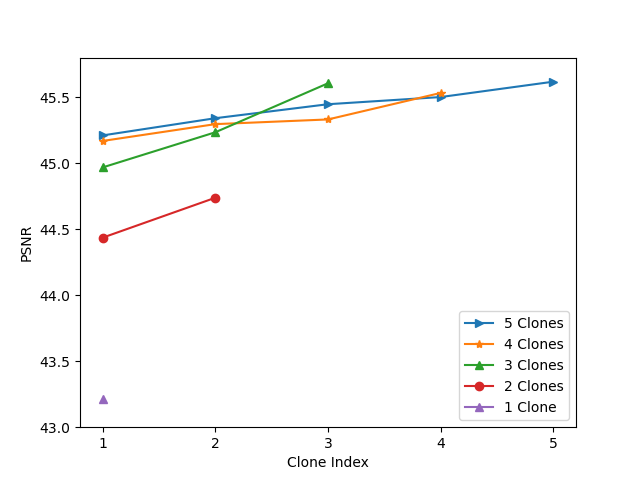}
		\label{fig_2_case}}
	\caption{Effect of (a) number of CNN layers and (b) number of clones on the testing PSNR performance of the PCN model. }
	\label{fig:7}	
\end{figure}

Fig. \ref{fig:7}(a) shows the effect of the number of CNN layers on the image PSNR of the PCN model applied on the testing dataset. The number of clones was fixed at 4. 
The result suggests a 10-layer RED-CNN can work well for the PCN model. Adding more layers did not improve the performance significantly. The result is consistent with \cite{b28}.

Fig. \ref{fig:7}(b) shows the effect of the number of clones on the PCN model performance. The PSNR of the output image of each clone  was plotted versus the clone index in each PCN model. The total number of clones was varied from 1 to 5. The number of CNN layers was set to 10 based on the result from Fig. \ref{fig:7}(a). The curves show that the PSNR of the earlier clones in a PCN model was improved as the number of clones increased.  The PSNR of the last clone in each PCN reached its maximum when the number of clones was 3. The differences in peak PSNR were minor  among the PCN models with 3 clones, 4 clones, and 5 clones. 

Table II compares the choices of the basic CNN module for quantitative image quality evaluation of the PCN. Two basic models were compared, including CPCE \cite{b29} 
and RED-CNN \cite{b28}. The number of CNN layers was set to 10 in each comparing model and the number of clones was set to 4. The result suggests the RED-CNN was better than CPCE to serve as the basic CNN module for the parallel-clone network.

\subsection{Comparison with Other DL Models}

Table III summarizes the results of PSNR, SSIM, and RMSE for denoising the testing dataset using five different models: the proposed PCN, the original DnCNN, RED-CNN, and MAP-CNN. The PCN was implemented with 3 clones. The result of MAP-NN trained using the MSE loss was also included as MAP-NN$_\mathrm{MSE}$ in the study. Among the different models, the proposed PCN achieved the best quality as assessed by all the three metrics. Compared to DnCNN and RED-CNN which are equivalent to a PCN with single clone, the use of multiple clones in the proposed PCN led to a significant improvement.
The comparison of PCN with the sequential-clone models (MAP-NN$_\mathrm{MSE}$ and MAP-NN) indicates the parallel structure of PCN can be superior to the sequential structure. Note that the improvement of MAP-NN over MAP-NN$_\mathrm{MSE}$ was mainly from the use of a more advanced loss function in the former. This suggests that a combination of advanced loss functions with PCN would be able to further improve the performance of the PCN model, which will be explored in our future work.

\begin{table}[t]
	\centering 
	\caption{~Comparison of PCN with different types of the basic CNN module.}
	\begin{tabular*}{3.5in}{@{\extracolsep{\fill}}lccc}
		\hline
		Module	&PSNR &SSIM&RMSE \\
		\hline
		CPCE& 44.6365$\pm$1.1332& 0.9782$\pm$0.0050\ &0.0056$\pm$0.0007\\
		RED-CNN& \textbf{45.4235$\pm$1.1394}&\textbf{0.9846$\pm$0.0047}&\textbf{0.0054$\pm$0.0006}\\
		\hline
	\end{tabular*}
	\label{tab:my_label}
\end{table}

\begin{table}[t]
	\centering 
	\caption{~Comparison of the proposed PCN model with other DL models for LDCT image denoising.}
	\begin{tabular*}{3.5in}{@{\extracolsep{\fill}}lccc}
		\hline
		Method	&PSNR &SSIM&RMSE \\
		\hline
		DnCNN \cite{b35}& 44.1305$\pm$1.2569&0.9739$\pm$0.0057&0.0057$\pm$0.0007\\
		RED-CNN \cite{b28}& 44.5238$\pm$1.1924 & 0.9756$\pm$0.0055&0.0055$\pm$0.0006\\
		MAP-NN \cite{b29}& 45.4612$\pm$1.3908& 0.9829$\pm$0.0054\ &0.0054$\pm$0.0007\\
		MAP-NN$_\mathrm{MSE}$& 43.3698$\pm$1.1724& 0.9612$\pm$0.0055\ &0.0070$\pm$0.0011\\
		Proposed (PCN) & \textbf{45.7775$\pm$1.1057}&\textbf{0.9855$\pm$0.0045}&\textbf{0.0053$\pm$0.0006}\\
		\hline
	\end{tabular*}
	\label{tab:my_label}
\end{table}

\begin{figure}[t]
	\centering
	\includegraphics[trim=0cm 0cm 0cm 1.55cm, clip,width=3.2in]{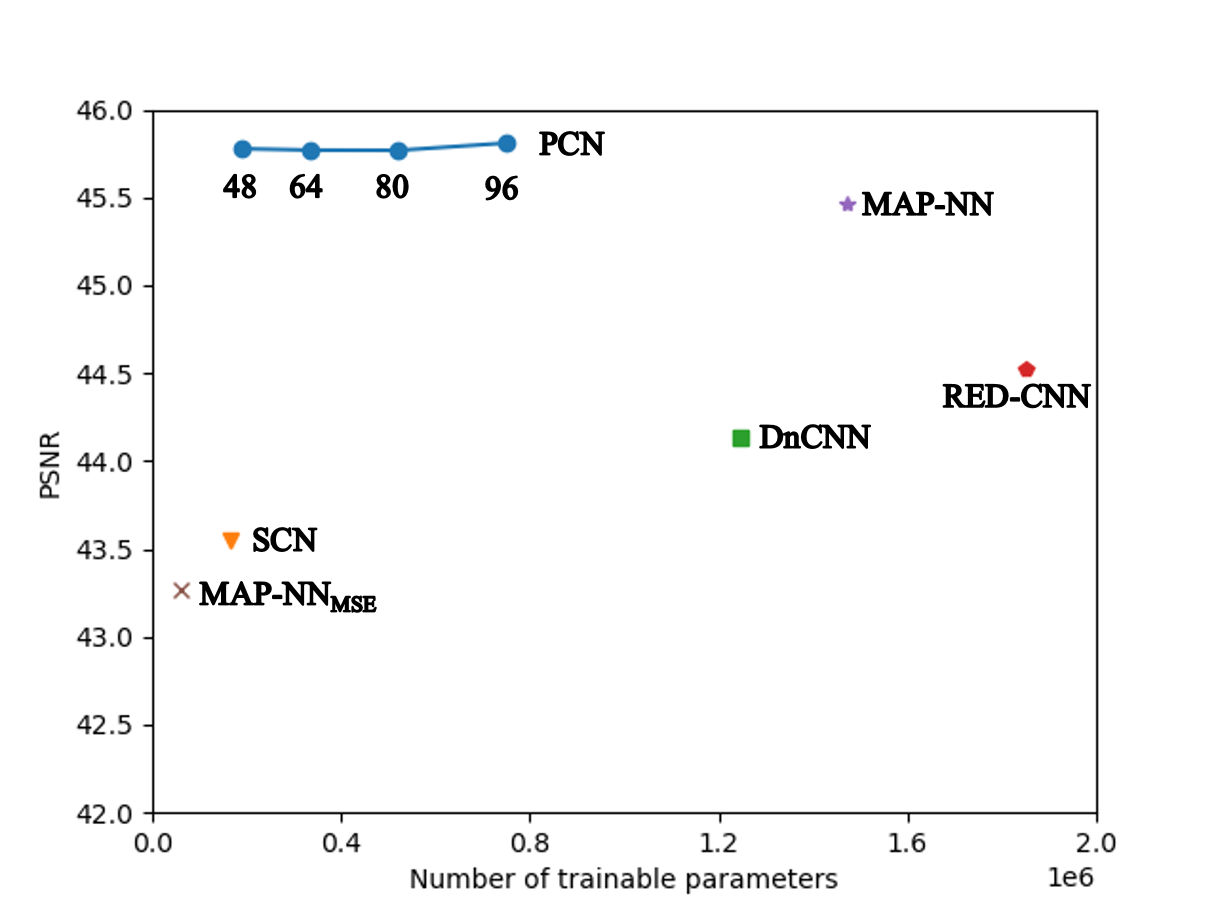}
	\caption{~Comparison of the PSNR and model complexity of different models.}
	\label{fig:9}	
\end{figure}

Different DL models have different model complexities. Fig. \ref{fig:9} shows the PSNR achieved by each DL model versus the number of trainable parameters) in the model. In addition to the use of 48 convolutional kernels, we also include the result for the use of 64, 80, and 96 kernels in the PCN. The result indicates the increased number of kernels has a minimal effect on PSNR once it exceeds 48. The baseline SCN was composed of four RED-CNN clones, but its performance was worse than the original RED-CNN \cite{b28}, mainly because  the former used a smaller kernel size and a much less number of kernels. Compared to the MAP-NN$_\mathrm{MSE}$, the use of advanced loss functions in the MAP-NN improved PSNR but also largely increased the model complexity. In comparison, the PCN achieved better PSNR performance with fewer parameters.

\begin{figure*}[t]
	\centering
	\includegraphics[width=5.5in]{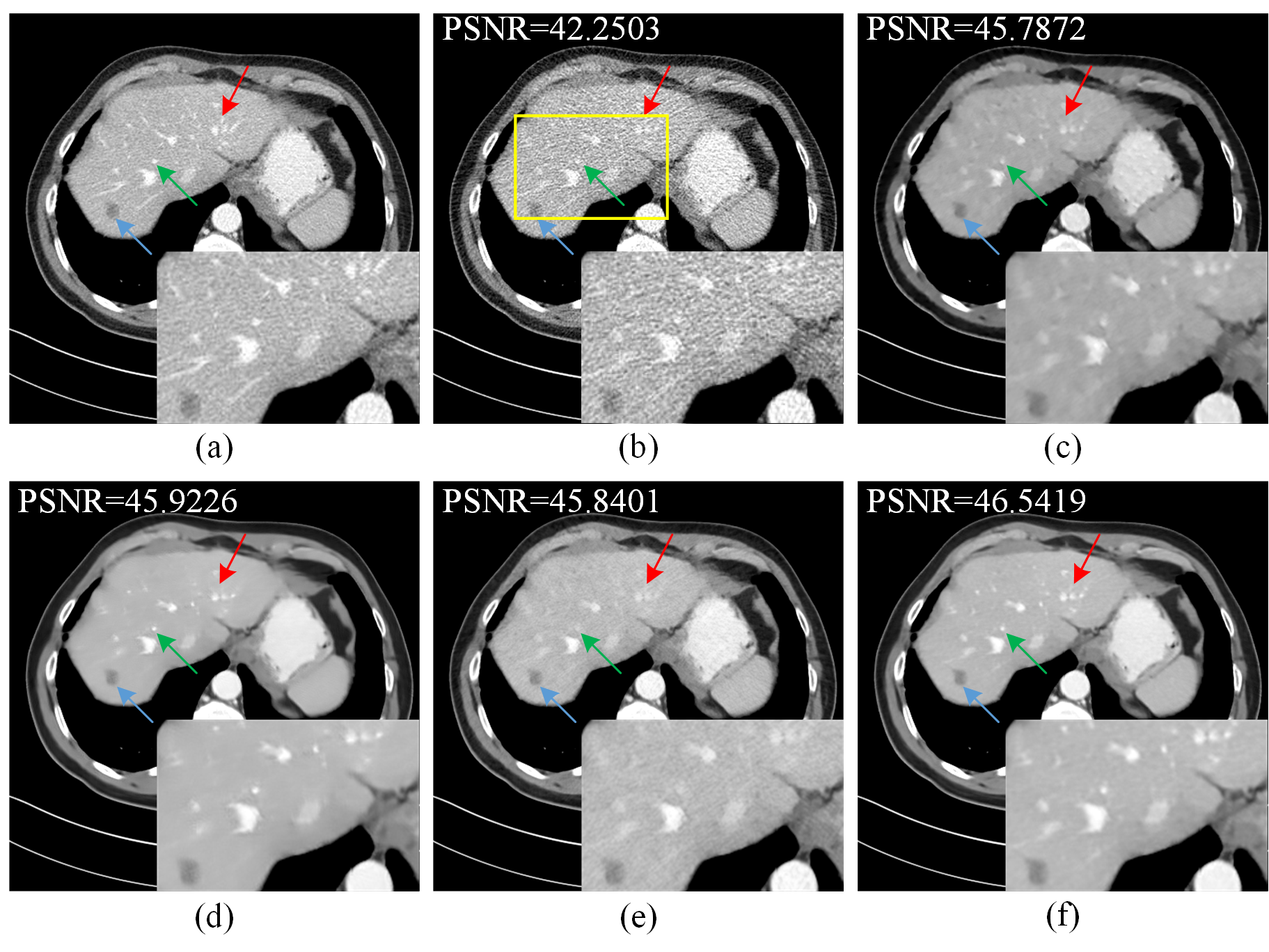}
	\caption{Comparison of the images denoised by different deep-learning models. (a) Normal-dose CT, (b) LDCT, (c) DnCNN, (d) RED-CNN, (e) MAP-NN. (f) Proposed PCN. }
	\label{fig:10}	
\end{figure*}

Fig. \ref{fig:10} shows the denoised images obtained by different models. The DnCNN and RED-CNN generally oversmoothed the liver background. The MAP-NN had a closer image appearance to to the normal-dose CT reference image, but some details were lost or with lower contrast, as pointed by the arrows. In comparison, the PCN provided generally higher image quality and better visual quality. 

These results together indicate the proposed PCN model can outperform existing DL models for LDCT image denoising.

\section{Conclusion}
In this paper, we have developed a simple yet efficient parallel-clone network architecture for LDCT image denoising. The model uses modularized clones with shared weights and exploits the benefits of parallel input, parallel-out loss, and clone-to-clone feature transfer. It has a similar or less number of unknown parameters as compared to conventional deep learning models but can significantly improve the learning process. Experimental results using the Mayo LDCT dataset demonstrated the improvement of the proposed parallel-clone network model over conventional sequential models.

\section*{Acknowledgment}

The authors thank Dr. Cynthia H. McCollough and the Mayo team for providing the LDCT dataset used in this study.

\end{document}